\documentstyle[twoside,fleqn,espcrc2]{article}

\newcommand{\AmS}{{\protect\the\textfont2
  A\kern-.1667em\lower.5ex\hbox{M}\kern-.125emS}}

\title{Finite temperature $QED_3$ with light fermions
\thanks{Talk given at the conference LATTICE'92}}

\author{J.B. Kogut and J.-F. Laga\"e\\
\vspace{0.25cm}
Department of Physics, University of Illinois at Urbana-Champaign,
Urbana IL 61801,USA}

\begin{document}

\begin{abstract}
Non-compact $QED_3$ is simulated both in the quenched and unquenched cases.
In particular, we investigate the restoration of chiral symmetry at finite
temperature. We also compute the zero temperature spectrum of the theory,
including (in the quenched case) the dynamical fermion mass. From these two
set of data, one can obtain estimates for the ratio of the mass gap to the
critical temperature, of particular interest for applications to high$-T_c$
superconductivity.
\end{abstract}

\maketitle

\section{Introduction}
Over the last few years, $QED_3$ has attracted a lot of attention because of
potential applications to models of high$-T_c$ superconductivity. However,
at the present time, the phase diagram of the theory is still not known
precisely. There have been several analytical studies but approximations
are often drastic and different results are obtained depending on what
assumptions are made. Hence the interest of trying to see if a lattice
gauge theory simulation can sort out this situation.

To avoid any confusion, it is worth mentioning that we are considering here
the 4-component formulation of $QED_3$ where the Dirac algebra is represented
by the $4\times4$ matrices $\gamma_0$, $\gamma_1$ and $\gamma_2$.
This formulation  preserves parity
and possesses an SU(2) group of chiral symmetries generated by $\gamma_3$,
$\gamma_5$ and $\gamma_3\gamma_5$. It also arises naturally as
the continuum limit of the staggered fermion construction in (2+1) dimensions
\cite{BURDEN}. If a mass term is
spontaneously generated, the above SU(2) symmetry will break down to the U(1)
generated by $\gamma_3\gamma_5$ and there will appear two Goldstone
bosons associated with the operators $\bar{\psi}\gamma_3\psi$ and
$\bar{\psi}\gamma_5\psi$.

Before discusssing the numerical simulation, it is useful to make a brief
review of the existing analytical results. These are all based on approximate
treatments of the Schwinger-Dyson equation for the fermion self-energy. The
first approximation which was considered is the 1/$N_f$ expansion ($e^2N_f$
fixed) \cite{APPELQUIST}
where $N_f$ is the number of fermion flavors. To first order, there
is no vertex correction and the photon propagator becomes:
\begin{equation}
D(p)={1/[p^2 + {e^2N_f \over 8}p}]
\end{equation}
A numerical solution of the Schwinger-Dyson equation then gives for the
critical number of flavors beyond which chiral symmetry is restored:
$N_c = 3.28$. This result has been criticized by Pennington and Walsh
\cite{PENNINGTON},
who by considering a plausible expression for the interaction vertex arrive
at the conclusion that chiral symmetry is always broken (however high $N_f$
can be).The finite temperature theory has also been investigated \cite{DOREY}.
In this case, the longitudinal photons are screened, but the transverse photons
remain unscreened:
\begin{equation}
D_{00} = {1 \over q^2+M^2} \qquad D_{ij} = {1 \over q^2}
\end{equation}
The transverse photons therefore give rise to severe infrared divergences
and it is not known at present how this problem can be cured. Dorey and
Mavromatos have assumed that transverse photons can be neglected in the
Schwinger-Dyson equation and have computed the critical temperature in this
case. They obtain in this way for $N_f=2$ a ratio of the mass gap to the
critical temperature which is of order 10 (much bigger than in BCS-like
theories, hence the potential interest for models of high-$T_c$
superconductivity).
Because of all the uncertainties in the above mentioned results, it would be
very interesting to have lattice measurements of the relevant quantities.
We will first discuss the techniques that we have at our disposal to carry out
such a project and will then present our current results for the chiral
condensate (both at zero and non-zero temperature) and for the meson spectrum.

\section{Symmetry Breaking on the lattice}
When studying chiral symmetry on the lattice, we encounter from the start a
fundamental problem: since the number of degrees of freedom is finite, a
symmetry can not be spontaneously broken. Therefore
$\langle\bar{\psi}\psi\rangle$ will
tend to zero when $m \rightarrow 0$. However, one can hope that there will
be two regimes in the dependence of $\langle\bar{\psi}\psi\rangle$ on m.
In the intermediate mass range, $\langle\bar{\psi}\psi\rangle$
would be essentially independent of the size of
the lattice, whereas for small masses, finite-size effects would show up and
$\langle\bar{\psi}\psi\rangle$ would drop sharply to zero.
If this is the case, then we can obtain an estimate of the infinite volume
value of $\langle\bar{\psi}\psi\rangle$
by extrapolating from the intermediate mass range. Although this technique
usually provides useful information, it may not be powerful enough to really
identify a phase transition. However, further checks are available to confirm
the presence or the absence of a phase transition. To be sure that chiral
symmetry is really broken in the continuum, one has to show that
$\langle\bar{\psi}\psi\rangle$ obeys the required scaling behaviour with
respect to $\beta$.
In 3 dimensions $\beta^2\langle\bar{\psi}\psi\rangle$ (= continuum value of
$\langle\bar{\psi}\psi\rangle$ in units of $e^4$) has to be constant. In
practice, one is looking for a range of $\beta$ where there is a plateau in
$\beta^2\langle\bar{\psi}\psi\rangle$. Note however that finite size effects
can significantly delay the appearance of such a plateau. In \cite{HANDS}
for example
it was shown that chiral symmetry is indeed broken in quenched QED3 but
lattices
of size ranging up to $(80)^3$ were necessary for that purpose. In the case
where the extrapolation method seems to indicate the presence of a phase
transition at finite $\beta$, we can assess its existence by using techniques
borrowed
from the ``theory'' of critical phenomena. At the critical value of the
coupling, for example, it is expected that $\langle\bar{\psi}\psi\rangle$
will behave like $m^{1/\delta}$ (where $\delta$ is one of the critical
exponents
characterizing the transition).

Independently of these techniques based on measurements of the chiral
condensate, one can also gain information about the phase diagram by
considering the meson spectrum. One has the following alternative: Either
there are parity doublets in the limit of massless fermions (and the $\pi$
and the $\sigma$ would be degenerate for example) or this is not the case and
the mass of the pion satisfies the usual PCAC relation. Consequently for m=0,
the ratio $M_{\pi}/M_{\sigma}$ should be a step function centered at
$\beta = \beta_c$. If m is not exactly zero then the function will be smoothed.
This has recently been given a quantitative form by applying the hypothesis
of correlation length scaling \cite{KOCIC}.

\section{The chiral condensate at T=0}

We have extended the results of \cite{DAGOTTO} by measuring the chiral
condensate on a
$(12)^3$ lattice. Put together, these results indicate that finite size effects
are strong, which makes the application of the techniques mentioned above
rather difficult. But, if we try anyway, the data presented in fig.1 amd 2
seem to indicate the presence of a phase transition around $\beta=0.2$ for
$N_f=4$ and around $\beta=0.26$ for $N_f=3$ (although the later is less
compelling because several curves are compatible with linear behaviour).
The critical exponents would be respectively $\delta=3.7$ and 2.7 but the
uncertainties are large. Clearly these results require some further
confirmation either from measurements on bigger lattices or from some
independent determination of $\delta$.

\section{Spectrum computations}

We used Kogut-Susskind fermions and computed the correlators between mesonic
operators
which are local in the lattice variables.
Because of lack of space, only the results for the pion will be presented here.
We have accumulated data for $N_f=0$, 2 and 4 at values of the coupling ranging
from $\beta=0.2$ to $\beta=0.4$ and for masses m=0.0125, 0.025 and 0.05.
In the quenched case, chiral
symmetry is clearly broken: we observe the behaviour expected from PCAC over
a wide range of values of $\beta$ (Fig.3) and the finite-size effects can be
investigated systematically. For $N_f=2$ (Fig.4), at $\beta=0.2$, we observe
on a $(12)^2\times24$ lattice that chiral symmetry is broken. At $\beta=0.4$,
strong
finite-size effects are present but a comparison of our $(12)^2\times24$ and
$(14)^2\times24$ results suggest that chiral symmetry will still be broken in
the infinite volume limit. For $N_f=4$ (Fig.5), chiral symmetry is broken at
$\beta=0.2$ but probably not at $\beta=0.4$ although finite size effects
remain important even on a $(16)^2\times24$ lattice.
If the above results are correct, they would confirm the analytical
expectations based on a $1/N_f$ expansions, namely that the critical value of
$N_f$ beyond which chiral symmetry is restored is around 3. At this point,
it should be mentioned that in their lattice investigation of compact
and non-compact QED3, Azcoiti and Luo \cite{AZCOITI} arrive at a rather
different picture of the phase diagram than the one suggested here.

\section{The chiral condensate at $\bf{T\not=0}$}

Our measurements of $\langle\bar{\psi}\psi\rangle$ at $N_f=2$ on
$(12)^2\times4$
and $(8)^2\times4$ lattices are shown on Fig.6. Using the extrapolation method
mentioned previously, we would conclude that chiral symmetry is broken at
$\beta=0.2$ (Note that the extrapolated value of $\langle\bar{\psi}\psi\rangle$
increases with the spatial size of the lattice) whereas it is not broken for
$\beta\geq0.3$. However, one has to be very careful at this point.
Indeed, if this is really indicative of a transition in the
continuum, then we should see again the transition on lattices with $N_\tau=8$
and it should happen between $\beta=0.4$ and $\beta=0.6$. But nothing like
this is observed on Fig.7, instead all the curves extrapolate smoothly to
zero. In order to further test for the presence of a transition, we can plot
$\ln\langle\bar{\psi}\psi\rangle$ versus $\ln m$ (Fig.8) and see if there
is a curve which exhibits critical behaviour. Only $\beta=0.3$ would be
compatible with a straight line provided that we eliminate the point at
$m=0.00625$. The presence of this point in fact seems to indicate that
$\beta=0.3$ is an upper bound for the value of the critical temperature.
However, one should consider the possibility that this point is sensitive to
finite-size effects (although for $\beta\geq0.4$ further runs on
$(24)^2\times8$
lattices indicated that finite-size effects are small) and it is tempting
to speculate that the actual transition is not far from $\beta=0.3$. If
this were the case, the ratio of the critical temperature to the square of
the electric charge would be close to $T/e^2=0.0375$, almost an order of
magnitude bigger than the result of Dorey and Mavromatos (0.0043)!
Further runs at lower values of $\beta$ are needed in order to resolve the
ambiguity. Ideally one would like to obtain a curve with positive curvature
on Fig.8 since it would give us a lower bound on the value of the critical
temperature. Finally, measurements of screening length (analogous to meson
spectrum calculations at zero temperature) could also provide useful
information for locating the position of the phase transition. We hope to
be able to report on this in the near future.

\section{Conclusion}
Our results seem to give some support to the following two statements:
\begin{enumerate}
\item At zero temperature, chiral symmetry is restored beyond some critical
value ($N_c$) of the number of flavors and $N_c$ is close to 3.
\item For $N_f=2$, the temperature at which chiral symmetry is restored is
much higher than the current analytical prediction \cite{DOREY}.
\end{enumerate}
However, as is clear from the discussion given in the text, further work
is still needed in order to completely clarify the situation.

\vfill\eject
\end{document}